\documentclass[pre,showpacs,amsmath,amssymb,two column,superscriptaddress]{revtex4}
\usepackage{amsmath}
\usepackage{amssymb}
\usepackage{amsfonts}
\usepackage{graphicx}
\usepackage{amsthm}

\newcommand{\R}{{\mathbb R}}
\newcommand{\RQ}{{\mathbb R}\setminus{\mathbb Q}}

\newcommand{\DTQW}{{DTQW}}

\newcommand{\ket}[1]{|{#1}\rangle}
\newcommand{\kb}[1]{|{#1}\rangle \langle{#1}|}

\newcommand{\bra}[1]{\langle{#1}|}

\newcommand{\psiCW}{{\varphi}}
\def\hsymbu#1{\smash{\lower1.7ex\hbox{\huge$#1$}}}
\def\hsymbl#1{\smash{\hbox{\huge$#1$}}}

\makeatletter

\newtheorem{thm}{Theorem}
\newtheorem{lem}{Lemma}
\newtheorem{col}{Corollary}
\newtheorem{remark}{Remark}
\newtheorem{conj}{Conjecture}

\begin{document}\title{Localization and Fractality in Inhomogeneous Quantum Walks with Self-Duality}
 \author{Yutaka Shikano}
 \email{shikano@mit.edu}
 \affiliation{Department of Physics, Tokyo Institute of Technology, Meguro, Tokyo, 152-8551, Japan}
 \affiliation{Department of Mechanical Engineering, Massachusetts Institute of Technology, Cambridge, MA 02139, USA}
 \author{Hosho Katsura}
 \email{katsura@kitp.ucsb.edu}
 \affiliation{Kavli Institute for Theoretical Physics, University of California Santa Barbara, CA, 93106, USA}
 \date{\today}
\begin{abstract}
We introduce and study a class of discrete-time quantum walks on a  one-dimensional lattice. 
In contrast to the standard homogeneous quantum walks, coin operators are inhomogeneous and depend on their positions in this class of models. 
The models are shown to be self-dual with respect to the Fourier transform, 
which is analogous to the Aubry-Andr\'e model describing the one-dimensional tight-binding model with a quasi-periodic potential. 
When the period of coin operators is incommensurate to the lattice spacing,  
we rigorously show that the limit distribution 
of the quantum walk is localized at the origin. 
We also numerically study the eigenvalues of the one-step time evolution 
operator and find the Hofstadter butterfly spectrum which 
indicates the fractal nature of this class of quantum walks. 
\end{abstract}
\maketitle
\section{Introduction}
A quantum walk (QW)~\cite{Gudder,Aharonov,Meyer}, 
the quantum mechanical analogue of the classical random walk (RW), 
is a useful tool to promote various fields. 
For instance, the QW is an important primitive in
universal quantum computation~\cite{Childs, Lovett}
and efficient quantum algorithms~\cite{Shenvi,Ambainis2,Buhrman,Magniez,Magniez2}. 
The realization of topological phases~\cite{Kitagawa} and 
quantum phase transitions~\cite{Chandrashekar} in optical lattice systems by the QW have been discussed. 
The relation between the Landau-Zener transition and the QW was found in Ref.~\cite{Oka}. 
The QW provides a new clue to more fundamental problems such as 
the quantum foundations of time~\cite{Shikano} and 
the photosynthesis and efficient energy transfer 
in biomolecules~\cite{Engel, Mohseni}. 
There are several important properties of the QW: 
(i) the inverted-bell like limit distribution, (ii) the quadratic speed-up of the variance to the classical RW, and (iii) the localization 
due to the quantum coherence and interference. 
The physical, mathematical, and computational properties of the 
QW are summarized in Refs. \cite{Kempe,KonnoRev,Kendon,Venegas-Andraca}. 

In this paper, we focus on the localization property of the QW 
and study a class of 
the discrete-time quantum walk (\DTQW) on a one-dimensional lattice with spatially inhomogeneous coins.
Throughout this paper, we define 
the localization such that the limit distribution of the \DTQW~divided by some power of the time variable 
has the probability density given by the Dirac delta function. 
The localization in the \DTQW~with homogeneous coins has so far been extensively studied. 
In the one dimensional \DTQW, the localization was shown in
the models with a three-dimensional coin~\cite{Inui}, a four-dimensional coin~\cite{Inui2}, a two-dimensional coin with memory~\cite{McGettrick},
a two-dimensional coin in a random environment~\cite{Joye} 
and a time-dependent two-dimensional coin; the Fibonacci QW~\cite{Romanelli}. 
On the other hand, the nature of the localization in 
the two-dimensional \DTQW~has been studied 
numerically~\cite{Mackay, Tregenna} and 
analytically~\cite{Inui3, Watabe}. 
The \DTQW~on the Cayley tree with a multi-state coin was also studied~\cite{Chisaki}. 
Furthermore, the recurrence properties of the \DTQW, 
i.e., the decay rate of the probability at the origin, were studied in Refs. \cite{Wojcik,Stefanak,Stefanak2,Chandrashekar2,Xu}. 
Recently, the localization property of the \DTQW~with the inhomogeneous two-dimensional coin has been 
analytically shown in the model 
in which the coin at the origin is different from the rest~\cite{Konno1,Konno2} 
and in the model of the periodically inhomogeneous coins~\cite{Linden}.
In this paper, we generalize the model investigated in Ref.~\cite{Linden} to the incommensurate case 
and show the localization property of the generalized model. 

The generalized model defined in the next section is inspired by 
the Aubry-Andr\'e model~\cite{Aubry_Andre}, which describes the one-dimensional tight-binding model with an incommensurate potential. 
The corresponding Schr\"odinger equation is called the almost Mathieu equation~\cite{Barry_Simon}: 
\begin{equation}
\psi(n+1)+\psi(n-1)+V \cos(2\pi \alpha n)\psi(n)=E\psi(n),
\label{Mathieu}
\end{equation}
where $\psi(n)$ is the amplitude of the wavefunction at the $n$th site, $\alpha$ is the irrational number, and $E$ is the eigenenergy. 
The model is often used to study the localization-delocalization transition~\cite{Sokoloff}. 
Aubry and Andr\'e have shown that all the eigenstates are localized when $V>2$ while they are extended when $V<2$. The model was shown to 
be self-dual with respect to the Fourier transform at $V=2$~\cite{Aubry_Andre}. 
The critical point $V=2$ is closely related to the Azbel-Hofstadter problem, which is the eigenvalue problem of the Bloch electron on a square 
lattice in a uniform magnetic field~\cite{Azbel, Hofstadter}. 
The energy spectrum shows a self-similar and fractal structure known as the Hofstadter butterfly~\cite{Hofstadter}. 
The multifractal property of the spectrum and wavefunctions have been studied in Refs.~\cite{Kohmoto_1983, Tang_Kohmoto} 
and its Bethe ansatz solution was 
discussed in Refs.~\cite{Wiegmann, Hatsugai_Kohmoto_Wu, Abanov}. 
In this paper, we show that our QW model is self-dual under 
the Aubry-Andr\'e duality transformation. We also study the eigenvalues of the one-step time evolution of our QW and find that the spectrum shows 
a self-similar and fractal structure in analogy to the Hofstadter butterfly.  

The paper is organized as follows. In Sec. \ref{AA_sec}, we define the \DTQW~with spatially inhomogeneous coins 
and show that this model is self-dual under a certain transformation, which is analogous to
the Aubry-Andr\'e duality. 
In Sec. \ref{Loc_sec}, 
we rigorously show that the limit distribution of our DTQW model is localized at the origin, 
which means that the distribution 
has the finite support for any time. In Sec. \ref{AA2_sec}, we numerically 
study the eigenvalues of the unitary operator for the one-step time evolution in our model and show the spectrum as a function of the 
inverse period (the Hofstadter butterfly diagram). 
Section \ref{con_sec} is devoted to the summary and discussion.
\section{Inhomogeneous Quantum Walk and Aubry-Andr\'e Duality} \label{AA_sec}
Let us define the one-dimensional \DTQW~\cite{Ambainis} as follows. 
First, we prepare the position and coin states denoted as $\ket{n} \otimes \ket{\zeta}= \ket{n,\zeta}$.  
Here, the position $n$ is along the infinite one-dimensional  
lattice $\mathbb{Z}$ (the integers) and 
$\ket{\zeta}$ is a normalized linear combination of 
$\ket{L} = ( 1 , 0 )^{{\bf T}}$ and $\ket{R} = ( 0 , 1 )^{{\bf T}}$, where ${\bf T}$ is the transposition. 
Second, we describe the time evolution of the \DTQW~using the unitary operator $U$ defined by two kinds of unitary operations ${\hat C}$ and $W$. 
The quantum coin flip ${\hat C} \in U(2)$ corresponds to the coin flip at each site in the RW. 
By the shift operator $W$, the position $n$ changes depending on the coin state:
\begin{align}
	W \ket{n,L} &:= \ket{n-1,L}, \notag \\
	W \ket{n,R} &:= \ket{n+1,R}.
\end{align} 
We can explicitly write the shift operator as  
\begin{equation}
W=\sum_n \left( |n-1,L \rangle \langle n,L| + |n+1,R \rangle \langle n,R| \right).
\label{shift}
\end{equation} 
Then, the one-step time evolution for the \DTQW~is expressed as $U = W (Id. \otimes \hat{C})$, where $Id.$ expresses the identity operator. 
Starting from the initial state $\ket{0,\phi}$, we apply $U$ to the state at every step. 
After the $t$ time steps, we obtain the probability distribution 
at the position $n$ as 
\begin{equation}
\Pr (n;t) = \sum_{\xi \in \{ L,R \} } \left| \bra{n,\xi} U^{t} \ket{0,\phi} \right|^2.
\end{equation}

\begin{figure}[ht]
\centering
	\includegraphics[width=8.5cm]{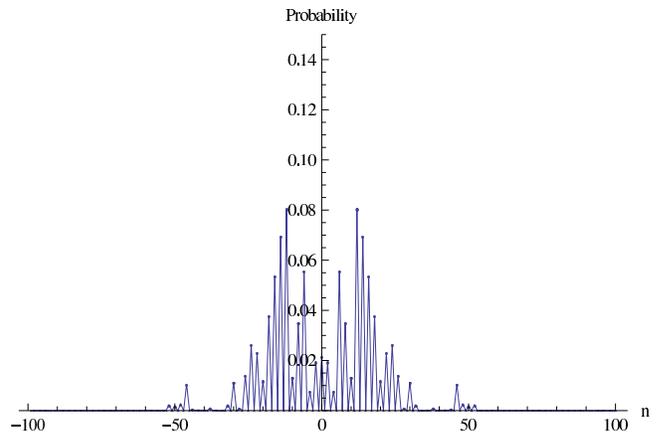}
\caption{Probability distribution in the inhomogeneous QW after the $1000$ time steps. Here, we take the initial coin state 
$(\ket{L}+\ket{R})/\sqrt{2}$ and the inverse period $\alpha = \pi/2$, which is an irrational number. 
Note that, the probability outside the plot range 
is zero within the numerical accuracy. 
}
\label{fig1}
\end{figure}

In the following, we introduce the \DTQW~with the spatially inhomogeneous coins. 
In this class of models, the quantum coin flip at each site can be different from the others. 
The coin operator $C$ acting on the position and coin states is then given by 
\begin{align}
C = & \sum_n \left[ 
  (\cos(2\pi \alpha n) |n,L \rangle + \sin(2\pi \alpha n) |n,R \rangle)\langle n,L| \right. \notag \\
 & \ \, \left. + (\cos(2\pi \alpha n) |n,R \rangle - \sin(2\pi \alpha n) |n,L \rangle)\langle n,R| \right] .
\label{coin}
\end{align}
More explicitly, the operator $C$ is written in the following form 
\begin{align}
C &= \sum_n \left[ \kb{n}
\otimes
\left( \begin{array}{cc}
\cos (2\pi \alpha n) & -\sin (2\pi \alpha n) \\
\sin (2\pi \alpha n) & \cos (2\pi \alpha n) 
\end{array}\right) 
\right] \notag \\
&:= \sum_n \left[ \kb{n} \otimes {\hat{C}}_n \right] ,
\label{coin2}
\end{align}
where $\alpha$ is a real number ($\alpha \in \mathbb{R}$) and corresponds to the inverse period of the coin operations. 
This inverse period can even be incommensurate to the underlying lattice. 
It reminds us of the Aubry-Andr\'e model~\cite{Aubry_Andre}. We call this model the inhomogeneous QW
throughout this paper. Note that, this model is a generalization of that in  Ref. \cite{Linden}. 
As an example, in Fig.~\ref{fig1} we show the behavior of the probability distribution at the $1000$th step obtained by numerical iteration. 

We shall now show that the inhomogeneous QW is {\it always} self-dual under a certain duality transformation.
Let us introduce a duality transformation, which corresponds to the spatial Fourier transform~\cite{Grimmett}, 
\begin{eqnarray}
|n,{\tilde L}\rangle = \sum_m \left[ \sin(2 \pi \alpha m n )|m,L\rangle + \cos(2 \pi \alpha m n ) |m,R\rangle \right] , \notag \\
|n,{\tilde R}\rangle= \sum_m \left[ \cos(2 \pi \alpha m n )|m,L\rangle + \sin(2 \pi \alpha m n ) |m,R\rangle \right].
\label{AA_tr}
\end{eqnarray}
It should be noted that these states are not normalized. 
Let us now see the action of the shift operator $W$ on these states. One can easily check the following relations
\begin{eqnarray}
W|n,{\tilde L}\rangle = \cos(2\pi \alpha n)|n,{\tilde L}\rangle + \sin(2\pi \alpha n)|n,{\tilde R} \rangle, \notag \\ 
W|n,{\tilde R}\rangle= \cos(2\pi \alpha n) |n,{\tilde R}\rangle -\sin(2\pi \alpha n) |n,{\tilde L} \rangle.
\end{eqnarray}
It is now obvious that the shift operator $W$ acts like a coin operator in the dual basis. 
Similarly, the action of the operator $C$ in the dual basis is given by
\begin{eqnarray}
C|n,{\tilde L} \rangle = |n-1,{\tilde L}\rangle, \notag \\
C|n,{\tilde R} \rangle = |n+1,{\tilde R}\rangle,
\end{eqnarray}
and it can be regarded as the shift operator in the dual basis. 
In this way, the duality transformation completely interchange the roles of $W$ and $C$. 
Therefore, the inhomogeneous QW 
is self-dual under the transformation (\ref{AA_tr}).
\section{Limit Localized Distribution} \label{Loc_sec}
In this section, we {\it analytically} show the localization in the inhomogeneous QW as the following theorem.
\begin{thm}
	For any irrational $\alpha \in \RQ$, 
	the limit distribution of the inhomogeneous QW divided by any power of the time variable  
	is localized at the origin:
	\begin{equation}
		\frac{X_t}{t^{\theta}} \Rightarrow I~~~~(t \to \infty),
		\label{limit}
	\end{equation}
	where $X_t$ is the random variable for the position at the $t$ step~\cite[Ch. 2]{FG}, 
	$\theta \ ( > 0)$ is an arbitrary parameter, and ``$\Rightarrow$" means convergence in distribution~\cite{comment}.
	Here, the limit distribution $I$ has the probability density function $f(x) = \delta (x)~~(x \in \R)$, 
	where $\delta ( \cdot )$ is the Dirac delta function. Note that, the limit distribution $I$ is independent 
	of the parameter $\theta$.
	\label{main_th}
\end{thm}
This theorem is the main result of this paper and means that the distribution of the inhomogeneous QW has a finite support 
for almost all $\alpha$. 
We first show the following lemma to prove the above theorem in the case of specific rational $\alpha$. 
\begin{lem}
	When $\alpha = \frac{P}{4Q} \in \mathbb{Q}$ with relatively prime $P$ (odd integer) and $Q$, the inhomogeneous QW 
	is restricted to the finite interval $[-Q,Q]$. 
	\label{lem1}
\end{lem}
\begin{proof}
    Let us express the state at the $t$ step evolving from the initial state $\ket{0,\phi}$ as
	\begin{equation}
		(WC)^{t} \ket{0, \phi} = \sum_{n \in \mathbb{Z} \atop \xi \in \{L,R\}} \psi_t ( n ; \xi ) \ket{n, \xi}.
	\end{equation}
	Then, we obtain the one-step time evolution of the coefficients $\psi_t (n; \xi)$ as follows: 	
	\begin{align}
		& \psi_{t+1} (n - 1 ; L ) \notag \\ & \ \ = \cos (2 \pi \alpha n) \psi_{t} ( n ; L ) - \sin (2 \pi \alpha n) \psi_{t} ( n ; R ), \notag \\
		& \psi_{t+1} (n + 1 ; R ) \notag \\ & \ \ = \sin (2 \pi \alpha n) \psi_{t} ( n ; L ) + \cos (2 \pi \alpha n) \psi_{t} ( n ; R ).
	\end{align}
	Since $\cos(\pm 2\pi \alpha mQ)=0$ for any odd integer $m$, that is, the diagonal elements of the coin operator are zero at position $\pm mQ$, 
	we immediately arrive at
	\begin{align}
		\psi_{t+1} ( \pm m Q - 1 ; L ) & = \pm (-1)^{\frac{mP+1}{2}} \psi_{t} ( \pm m Q ; R ), \notag \\
		\psi_{t+1} ( \pm m Q + 1 ; R ) & = \mp (-1)^{\frac{mP+1}{2}} \psi_{t} ( \pm m Q ; L ).
	\end{align}
The above equations mean that 
the quantum walker is reflected 
at the position $n = \pm m Q$ and cannot be transmitted across this point  
(see Fig.~\ref{fig}). 
Since the initial position is localized at the origin ($n = 0$), the quantum walker undergoes reflection only at the sites $\pm Q$. 
Therefore, we conclude that the distribution of the inhomogeneous QW has the finite support $[-Q,Q]$ for any time $t$.
\begin{figure}[ht]
\centering
	\includegraphics[width=8cm]{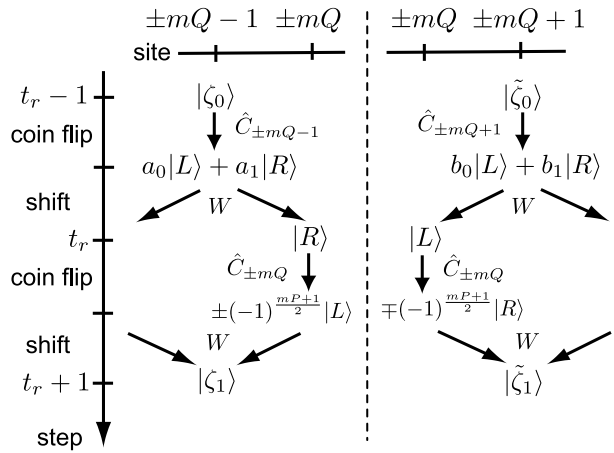}
\caption{
Time evolution of the inhomogeneous QW around the site $\pm m Q$ at which the quantum walker is reflected. 
The coin operator $C$ and shift operator $W$ act on the wavefunction $\ket{\psi_n (t)} = 
( \psi_t (n, L), \psi_t (n, R) )^{{\bf T}}$ at each site $n$. 
The wavefunction around the reflection point $\pm m Q$ is indicated 
from the $(t_r - 1)$th step to the $(t_r + 1)$th step.}
\label{fig}
\end{figure}
\end{proof}
From Lemma 1, we can conclude that our inhomogeneous QW is finitely restricted when $\alpha=\frac{P}{4Q} \in \mathbb{Q}$. 
We now try to extend this result to the case of $\alpha \in \RQ$ and show Theorem~\ref{main_th}. 
The following well-known theorem in number theory is useful. 
The proof of this theorem can be found in Ref.~\cite{Hardy}.
\begin{thm}[Dirichlet's Approximation Theorem{~\cite[Theorem 185]{Hardy}}]
	For any irrational number $\alpha \in \RQ$, there is an infinity of the fractions $a/b$ which satisfy  
	\begin{equation}
		\left| \alpha - \frac{a}{b} \right| < \frac{1}{b^2}.
	\end{equation}
	\label{dir}
\end{thm}
Theorem~\ref{dir} implies the following corollary which will be used to prove Theorem 1. 
\begin{col}
	For any irrational number $\alpha \in \RQ$, there is an infinity of the fractions $\frac{P}{4Q}$ with the relatively prime $P$ (odd integer) and 
	$Q$ which satisfy
	\begin{equation}
		\left| \alpha - \frac{P}{4Q} \right| < \frac{1}{4 Q^2}.
	\end{equation}
	\label{dir2}
\end{col}
\begin{proof}
	From Theorem~\ref{dir}, for any irrational number $\alpha^{\prime} \in \RQ$, 
	there is an infinity of the fractions $P^{\prime}/Q$ which satisfy
	\begin{equation}
		\left| \alpha^{\prime} - \frac{P^{\prime}}{Q} \right| < \frac{1}{Q^2}.
	\end{equation}
	$P^{\prime}$ can be expressed as $P^{\prime}= P \times 2^\ell$ with an odd integer $P$ and a non-negative integer $\ell$. 
	It is noted that $P$ and $Q$ are relatively prime 
	since $P^{\prime}/Q$ is an irreducible fraction. Therefore, dividing both sides of the above equation by $2^{\ell+2}$, we obtain
	\begin{equation}
		\left| \frac{\alpha^{\prime}}{2^{\ell+2}} - \frac{P}{4Q} \right| < \frac{1}{2^{\ell+2} Q^2} \leq \frac{1}{4 Q^2}.
	\end{equation}
	Since $\alpha := \alpha^{\prime}/2^{\ell+2}$ must be irrational, we obtain the desired result.
\end{proof}
In the following, we prove Theorem \ref{main_th}.
\begin{proof}[Proof of Theorem \ref{main_th}]
For any $\epsilon > 0$, there exist the relatively prime $P$ (odd integer) and $Q > \frac{\pi}{2 \epsilon}$ such that 
\begin{equation}
	\alpha = \frac{P}{4Q} + \delta,
	\label{alpha_con}
\end{equation}
where $-1/(4 Q^2) < \delta < 1/(4 Q^2)$ according to Corollary~\ref{dir2}. 
We calculate the diagonal elements of the coin operator at the $\pm Q$th site as
\begin{equation}
	\cos \left[ 2 \pi \left( \frac{P}{4Q} + \delta \right) (\pm Q) \right] = (-1)^{(P+1)/2} \sin (2 \pi Q \delta).
\end{equation}
Note that both the diagonal elements of ${\hat C}_{\pm Q}$ are equal. We obtain
\begin{equation}
	\left| (-1)^{(P+1)/2} \sin (2 \pi Q \delta) \right| < \frac{\pi}{2 Q} < \epsilon.
\end{equation}
Since $\epsilon$ can be arbitrarily small, the diagonal element is zero in the limit.
Then, along the same lines as Lemma~\ref{lem1}, the inhomogeneous QW is restricted to the finite interval 
$[-Q, Q]$. Since the time variable $t$ and the parameter $\epsilon$ are independent, 
the time variable $t$ and the parameter $Q$ are also independent.
Therefore, we obtain the desired result (\ref{limit}). 
\end{proof}
Analogous to the above discussion, we show the localization property of the \DTQW~with spatially random coins as follows. 
From Lemma~\ref{lem1}, we obtain the following remark, which was implicitly shown in Ref.~\cite{Linden}.
\begin{remark}
	If the diagonal elements of the coin operator at a position on each side the origin are zero, then the DTQW is finitely restricted, i.e., 
	the localization at the origin as in Eq. (\ref{limit}). \label{rem}
\end{remark}
Therefore, the limit distribution of the \DTQW~with a randomly chosen quantum coin ${\hat{C}}_{n, {\rm random}}$ with 
uniform distribution in $U(2)$ at each position $n \ ( \in {\mathbb Z} )$ 
divided by any power of the time variable is also localized. 
This is because the diagonal elements of ${\hat{C}}_{n, {\rm random}}$ must be zero at some positions. 
This result is relevant to Ref.~\cite{Joye}.
\section{Fractal Property} \label{AA2_sec}
In this section, we study the eigenvalues of the operator $WC$, which is the one-step time evolution in the inhomogeneous QW 
and is defined in Eqs. (\ref{shift}, \ref{coin}). 
We show the general properties of the eigenvalues as well as the numerically obtained distribution of them, which suggests the 
self-similar and fractal structure in the spectrum in analogy to the Hofstadter butterfly problem. 

We first show the following theorem for the eigenvalues of the operators $WC$ and $CW$. 
\begin{thm}
    All the eigenvalues of the operators $WC$ and $CW$ are identical including the degeneracy. 
	\label{eigen}
\end{thm}
\begin{proof}
Let $\ket{\mu}$ be an eigenvector of the operator $WC$ with a nonzero eigenvalue $\mu$: 
	\begin{equation}
		WC \ket{\mu} = \mu \ket{\mu}.
	\end{equation}
Note that, all the eigenvalues of $WC$ are nonzero and have modulus $1$ since $WC$ is a unitary operator. 
Multiplying $C$ from the left, one obtains 
	\begin{equation}
		CW (C\ket{\mu}) = \mu (C\ket{\mu}).
	\end{equation} 
Since $C$ is also a unitary operator, $C\ket{\mu}$ is a nonzero eigenvector of $CW$ with the eigenvalue $\mu$. 
To complete the proof, we need to repeat the same argument for $CW$. 
Let $\ket{\lambda}$ be an eigenvector of $CW$ with an eigenvalue $\lambda$: 
	\begin{equation}
		CW \ket{\lambda} = \lambda \ket{\lambda}.
	\end{equation}
Multiplying $W$ from the left, one obtains 
	\begin{equation}
		WC (W \ket{\lambda}) = \lambda (W \ket{\lambda}).
	\end{equation}
Since $W$ is a unitary operator, $W\ket{\lambda}$ is a nonzero eigenvector of $WC$ with the eigenvalue $\lambda$. 
The identical degeneracy of $WC$ and $CW$ follows from the fact that the unitary operator $C$ or $W$ preserves the orthogonality of the states with the same eigenvalue. 
\end{proof}

\begin{remark}
Theorem~\ref{eigen} holds not only for the inhomogeneous QW but also for the standard {\DTQW}, 
in which the coin operator does not depend on the positions. 
\end{remark}

According to Theorem~\ref{eigen}, we have only to study the eigenvalues of $CW$. 
In the following, we focus on the case of the rational number $\alpha = \frac{P}{4Q} \in \mathbb{Q}$. In this case, from Lemma~\ref{lem1}, 
we can express $CW$ as a $4Q \times 4Q$ irreducible matrix. 
Let us first derive a finite-dimensional matrix representation of $CW$. 
We can express the wavefunction at the $t$ step evolving from the state $\ket{0,{\tilde \phi}}$ by $CW$:
\begin{equation}
(CW)^t \ket{0,{\tilde \phi}}=\sum_{n \in \mathbb{Z} \atop \xi \in \{ L,R \} }\psiCW_t (n,\xi)\ket{n,\xi}.
\end{equation}
Along the same lines as the proof of Lemma~\ref{lem1}, one obtains the one-step time evolution of the coefficients $\psiCW_t (n, \xi)$ as follows:
\begin{equation}
\left(\begin{array}{c} \psiCW_{t+1}(n; L) \\ 
\psiCW_{t+1}(n; R) \end{array}\right)={\hat C}_n 
\left(\begin{array}{c} \psiCW_{t}(n+1; L) \\
\psiCW_t (n-1; R) \end{array}\right),
\label{eq: CW_dynamics}
\end{equation}
where ${\hat C}_n$ is defined in Eq. (\ref{coin2}).

Since the initial condition is $\ket{0,{\tilde \phi}}$, one can clearly see that $\psiCW_{t}(-Q; L)=\psiCW_{t}(Q; R)=0$ for any time $t$ 
and hence 
the inhomogeneous QW by $CW$ is also restricted to the finite interval $[-Q, Q]$ like Lemma~\ref{lem1}.
It is convenient to introduce the vector ${\vec{\psiCW}}_t$ 
defined by the coefficients $\psiCW_t (n; \xi)$ as
\begin{eqnarray}
{\vec{\psiCW}}_t &=& (\psiCW_t(-Q; R), \psiCW_t(-Q+1;L), \psiCW_t(-Q+1;R), \notag \\
&&..., \psiCW_t(Q-1; L), \psiCW_t (Q-1; R), \psiCW_t (Q; L))^{{\bf T}}.
\end{eqnarray}
Note that, ${\vec{\psiCW}}_t$ is the $4Q$-dimensional vector. 
Using ${\vec{\psiCW}}_t$, we can simply express Eq. (\ref{eq: CW_dynamics}) as 
\begin{equation}
{\vec{\psiCW}}_{t+1}={\sf CW}{\vec{\psiCW}}_t.
\end{equation}
Here, ${\sf C}$ and ${\sf W}$ are $4Q \times 4Q$ matrices and their explicit forms are given by
\begin{eqnarray}
{\sf C} &=& \left(\begin{array}{cccccc}
\hspace{-1.5mm}(-1)^{\frac{P+1}{2}} & & & & & \hsymbu{0} \\
   &\hspace{-1.8mm}{\hat C}_{-Q+1}\hspace{-1.5mm}& & & & \\
 & &\hspace{-1.5mm}{\hat C}_{-Q+2}\hspace{-1.8mm}& & & \\
 & & &\hspace{-1.mm}\ddots & & \\
 & & & &\hspace{-1mm}{\hat C}_{Q-1}\hspace{-1.8mm}& \\
\hsymbl{0} & & & & & (-1)^{\frac{P+1}{2}}\hspace{-1.5mm}
\end{array}\right) \label {sf C}\\
{\sf W} &=& \left(\begin{array}{cccccc}
0&{\hat e}^{{\bf T}}_1& & & & \hsymbu{0} \\
{\hat e}_2&{\hat O}& {\hat e}_1 & & \\
 & {\hat e}_2&{\hat O}&{\hat e}_1& & \\
 & &\ddots &\ddots &\ddots & \\
 & & &{\hat e}_2&{\hat O}& {\hat e}_1 \\
\hsymbl{0}& & & &{\hat e}^{{\bf T}}_2& 0
\end{array}\right), \label{sf W}
\end{eqnarray}
where 
\begin{equation}
{\hat O}=\left(\begin{array}{cc} 0&0 \\ 0&0 \end{array}\right),~~
{\hat e}_1 = \left(\begin{array}{c} 1 \\ 0 \end{array}\right),~~
{\hat e}_2 = \left(\begin{array}{c} 0 \\ 1 \end{array}\right).
\end{equation}
We note that empty entries in the matrices are zeros. 
One can easily confirm that ${\sf C}$ and ${\sf W}$ are unitary matrices. 

\begin{figure}[tb]
\centering
	\includegraphics[width=8.5cm]{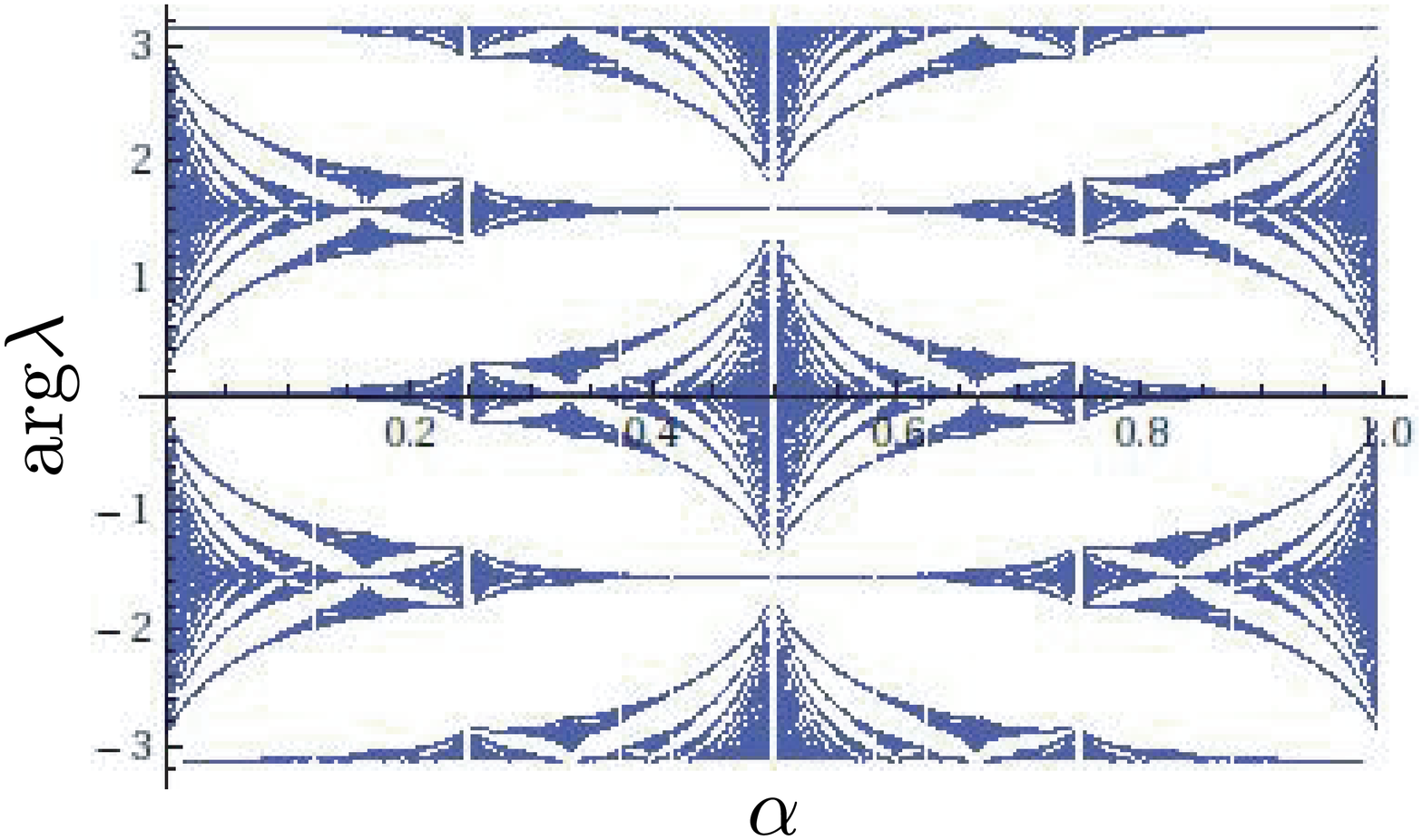}	
\caption{
The self-similar and fractal structure in the inhomogeneous QW.  
Arguments of the eigenvalues of $CW$ (vertical axis) are plotted as a function of the parameter $\alpha=\frac{P}{4Q}$ 
(horizontal axis) with $Q \le 50$. Note that, $P$ and $Q$ must be relatively prime. }
\label{fig2}
\end{figure}

We shall now study the eigenvalues of ${\sf CW}$. 
Since the matrix ${\sf CW}$ is unitary, the eigenvalues $\lambda (\in \mathbb{C})$ have modulus $1$, i.e., $|\lambda|=1$. 
Therefore, the argument of $\lambda$ faithfully represents the eigenvalues. 
Figure \ref{fig2} shows the spectrum as a function of the inverse period $\alpha$, obtained by exact diagonalization of ${\sf CW}$. 
Although we can only study the case of rational $\alpha= \frac{P}{4Q} \in \mathbb{Q}$ by numerical diagonalization, 
we can understand the spectrum of $CW$ with $\alpha \in \RQ$ as the irrational limit of a well-organized sequence of rational numbers, 
whose existence is guaranteed by Theorem~\ref{dir}.  
From Fig.~\ref{fig2}, one can clearly see the self-similar and fractal nature in the eigenvalue distribution. 
We also notice the following properties of the spectrum: 
\begin{description}
\item{(P1) All the eigenvalues at $\alpha$ are identical to those at $1-\alpha$. }
\item{(P2) The eigenvalues come in complex conjugate pairs: for every eigenvalue $\lambda$, there is an eigenvalue $\lambda^*$. }
\item{(P3) The eigenvalues come in chiral pairs: for every eigenvalue $\lambda$, there is an eigenvalue $-\lambda$.}
\item{(P4) All the eigenvalues are simple, i.e., nondegenerate.}
\item{(P5) There are four eigenvalues $\lambda=\pm 1, \pm i$ for any $\alpha=\frac{P}{4Q} \in \mathbb{Q}$.}
\end{description}

Let us explain the above properties in more detail. 
(P1) simply means that the spectrum in Fig. \ref{fig2} is symmetrical with respect to the line $\alpha=1/2$. 
This property can be shown as follows. 
Suppose that ${\vec{\psiCW}}$ is the eigenvector of ${\sf CW}$ at the inverse period $\alpha$ with the eigenvalue $\lambda$. 
Then, according to Eq. (\ref{eq: CW_dynamics}), one obtains
\begin{align}
\lambda \psiCW (-Q; R) &= (-1)^{\frac{P+1}{2}}\psiCW (-Q+1; L), \notag \\
\lambda \left(\begin{array}{c} \psiCW (n; L) \\ 
\psiCW (n; R) \end{array}\right) &= {\hat C}_n (\alpha)
\left(\begin{array}{c} \psiCW (n+1; L) \\ 
\psiCW (n-1; R) \end{array}\right), ( n \in (-Q,Q)) \notag \\
\lambda \psiCW (Q; L) &= (-1)^{\frac{P+1}{2}}\psiCW (Q-1; R),  
\label{eq: CW_dynamics_2}
\end{align}
where ${\hat C}_n$ is denoted as ${\hat C}_n (\alpha)$ to emphasize the argument $\alpha$.
We now consider the following transformation:
\begin{equation}
\psiCW (n; L) \to \psiCW (-n; R),~~~~\psiCW (n; R) \to \psiCW (-n; L). 
\label{eq: inversion}
\end{equation}
Under this transformation, Eq. (\ref{eq: CW_dynamics_2}) becomes
\begin{align}
\lambda \psiCW (-Q; R) &= (-1)^{\frac{P+1}{2}}\psiCW (-Q+1; L), \notag \\
\lambda \left(\begin{array}{c} \psiCW (n; L) \\ 
\psiCW (n; R) \end{array}\right) &= {\hat C}_n (-\alpha)
\left(\begin{array}{c} \psiCW (n+1; L) \\ 
\psiCW (n-1; R) \end{array}\right), ( n \in (-Q,Q)) \notag \\
\lambda \psiCW (Q; L) &= (-1)^{\frac{P+1}{2}}\psiCW (Q-1; R).  
\label{eq: CW_dynamics_3}
\end{align}
The above equations are equivalent to ${\sf CW}{\vec{\psiCW}}=\lambda{\vec{\psiCW}}$ at the inverse period $1-\alpha$. 
Therefore, the sets of eigenvalues at $\alpha$ and $1-\alpha$ are identical. 
Furthermore, one can construct the eigenvectors of ${\sf CW}$ at $1-\alpha$ from those at $\alpha$ using the transformation (\ref{eq: inversion}). 
(P2) implies that the spectrum is symmetrical with respect to the line ${\rm arg}(\lambda)=0$ in Fig.~\ref{fig2}. On can easily show (P2) by taking 
the complex conjugate of Eq. (\ref{eq: CW_dynamics_2}) and see that ${\vec{\psiCW}}^*$ is the eigenvector of ${\sf CW}$ with the eigenvalue $\lambda^*$. 
(P3) combined with (P2) gives rise to the symmetry with respect to the line ${\rm arg}(\lambda)=\pi/2$ or ${\rm arg} (\lambda)=-\pi/2$ in Fig.~\ref{fig2}.  
(P3) is reminiscent of the chiral or particle-hole symmetry in the original Hofstadter problem, i.e., the eigenenergies of the Hamiltonian come  in pairs 
$\pm E$. This property holds not only for our inhomogeneous QW but also for rather generic classes of the DTQW. 
To show (P3), we consider the local gauge transformation: $\psiCW (n; \xi) \to (-1)^n \psiCW (n; \xi)$. 
Under this transformation, ${\vec{\psiCW}}$ becomes the eigenvector of ${\sf CW}$ with the eigenvalue $-\lambda$ according 
to Eq. (\ref{eq: CW_dynamics_2}). This proves (P3). 

Properties (P4) and (P5) requires a more elaborate analysis. To show (P4), we first show that $\psiCW (-Q; R)\ne 0$ if ${\vec{\psiCW}}$ is an eigenvector of ${\sf CW}$. 
Suppose for the sake of contradiction that $\psiCW (-Q; R)=0$. Using Eq. (\ref{eq: CW_dynamics_2}), it follows that ${\vec{\psiCW}}={\vec 0}$. 
But this contradicts the fact that ${\vec{\psiCW}}$ is the eigenvector of ${\sf CW}$. Therefore, $\psiCW (-Q; R) \ne 0$. 
It should be noted that $2 \pi n \frac{P}{4Q} \ne \pi \ell/2$ ($\ell \in \mathbb{Z}$) for $-Q< n < Q$. 
We next suppose that ${\sf CW}$ has two eigenvectors ${\vec{\psiCW}}^{(1)}$ and ${\vec{\psiCW}}^{(2)}$ with the 
same eigenvalue $\lambda$. Then, we define the linear combination of them as
\begin{equation}
{\vec{\psiCW}}'={\psiCW}^{(2)}(-Q; R) {\vec{\psiCW}}^{(1)} - {\psiCW}^{(1)}(-Q; R) {\vec{\psiCW}}^{(2)}
\end{equation}
and find ${\vec{\psiCW}}'(-Q; R)=0$. 
The vector ${\vec{\psiCW}}'$ is also an eigenvector of ${\sf CW}$ and this  contradicts the previous statement. Therefore, all the eigenvalues of ${\sf CW}$ are simple. 
Let us finally show the last property (P5). 
We first note that the product of the eigenvalues $\lambda_k$ $(k=1, ..., 4Q)$ is
\begin{equation}
\prod^{4Q}_{k=1} \lambda_k =-1
\label{det CW}
\end{equation}
since $\det {\sf CW}=\det({\sf C}) \det({\sf W})=-1$. 
Here, we have used the fact that
$\det {\sf C}=1$ and $\det {\sf W}=-1$. 
This can be shown by an explicit calculation using Eqs. (\ref{sf C}) and (\ref{sf W}). Next, from (P2) and (P3), 
we see the eigenvalues $\lambda_k$ come in quadruplets $\lambda_k$, $\lambda^*_k$, $-\lambda_k$, $-\lambda^*_k$ if 
$\lambda_k$ is neither $\pm 1$ nor $\pm i$. 
Since $|\lambda_k|^2=1$ and Eq. (\ref{det CW}), there exists $k$ such that $\lambda_k=-1$. 
The multiplicity of $\lambda_k=-1$ should be $1$ due to (P4). 
Then, noting the fact that the number of eigenvalues $4Q$ is a multiple of $4$, we can conclude that $\lambda_k=\pm i$ should also be included as 
the eigenvalues of ${\sf CW}$. This proves the statement (P5). Note that, the properties of the chiral symmetry (P3) and the existence of 
eigenvalues $\lambda= \pm 1$ are relevant to Ref.~\cite{Kitagawa} discussed in the standard DTQW.
\section{Conclusion and Outlooks} \label{con_sec}
We have introduced the class of the inhomogeneous QW with self-duality inspired by the Aubry-Andr\'e model~\cite{Aubry_Andre}. 
Our main result is Theorem~\ref{main_th}. 
We have analytically shown that the limit distribution of the inhomogeneous QW divided by any power of time is localized at the origin 
when the inverse period of coins $\alpha$ is irrational number. 
We emphasize that our analytical solution for the localization is powerful since the round-off error is inevitable in numerical simulations. 
We have also numerically studied the eigenvalue distributions of the quantum walk operator $WC$ in the inhomogeneous QW 
with $\alpha =\frac{P}{4Q} \in \mathbb{Q}$. 
The obtained spectrum shows the self-similar and fractal structure similar to the Hofstadter butterfly. 

\begin{figure}[ht]
\centering
	\includegraphics[width=8.5cm]{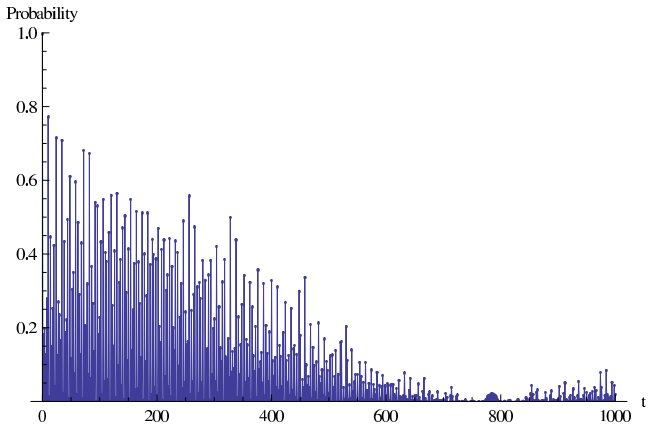}	
\caption{Recurrence property of the inhomogeneous QW with the initial coin state $(\ket{L}+\ket{R})/\sqrt{2}$ and the inverse period $\alpha = \pi/2$. 
The numerically obtained probabilities at the origin are shown up to $1000$ time steps. 
Note that the probability 
at the origin must be zero when the step $t$ is odd.}
\label{fig3}
\end{figure}

In our model, the following problems remain open: 
First, it might be possible to experimentally realize the inhomogeneous QW 
in quantum optics as pointed out in Ref.~\cite{Broome}. 
Second, the recurrence property of the inhomogeneous QW has not been analytically obtained. 
The example of the recurrence is numerically shown in Fig.~\ref{fig3}. 
Finally, we have not yet obtained the limit distribution in the case of the rational $\alpha \in \mathbb{Q}$ 
except for $\alpha = \frac{P}{4Q}$ with relative prime $P$ (odd integer) and $Q$. 
According to 
Ref.~\cite{Cantero}, the limit distribution of the DTQW is partly related to the eigenvalues of the quantum walk operator $WC$. From Fig.~\ref{fig2}, the 
eigenvalues in the considering case seem to be continuous. Therefore, we propose the following conjecture as a general property of the DTQW.
\begin{conj}
	The distribution of the DTQW is not localized if the quantum walk operator $WC$ only has continuous spectra and does not have embedded eigenvalues.
\end{conj}
This conjecture is motivated by Ref.~\cite{Simon} to show the property of the bound states from the spectrum of the discrete Schr\"{o}dinger operator, 
which includes the almost Mathieu equation (\ref{Mathieu}). 
Let us examine this conjecture for some cases in our inhomogeneous QW.  
It is easy to check that, in the case of $\alpha = P / 2 \ (P \in \mathbb{Z})$, $WC$ has a continuous spectrum and localization does not occur. 
In this case, the coin state is unchanged up to the phase factor since the coin operator is always $\left( \begin{array}{cc} 1 & 0 \\ 0 & 1 \end{array}\right)$ or 
$\left( \begin{array}{cc} -1 & 0 \\ 0 & -1 \end{array}\right)$. Therefore, the quantum walker does not undergo reflection at any point. 
Furthermore, in the case of $\alpha = 1/6$, Linden and Sharam showed that the standard deviation is proportional to the time variable $t$ 
for a sufficient large $t$~\cite{Linden}. These facts partially support the above conjecture.
\section*{Acknowledgment}
One of the authors (Y.S.) acknowledges Takuya Machida, Norio Konno, Etsuo Segawa, and Seth Lloyd for encouraging discussion. 
Y.S. is supported by JSPS Research Fellowships for Young Scientists (Grant No. 21008624). 
H.K. is supported by the JSPS Postdoctoral Fellowship for Research Abroad.


\begin{thebibliography}{99}
\bibitem{comment}
	Convergence in distribution is also used in the central limit theorem; 
	the discrete distribution can be approximated by the continuous distribution. 
	Therefore, the domain of the probability density function $f(x)$ for the limit distribution $I$ 
	is the real number $\R$. Please see Ref.~\cite[Part 3]{FG} for the formal definitions.
\end{thebibliography}
\end{document}